\documentclass[letterpaper,journal]{IEEEtran}
\ifdefined\pdfminorversion\pdfminorversion=7\fi
\usepackage{amsmath,amsfonts,amssymb,mathtools}
\usepackage{algorithmic}
\usepackage{algorithm}
\usepackage{array}
\usepackage[caption=false,font=normalsize,labelfont=sf,textfont=sf]{subfig}
\usepackage{textcomp}
\usepackage{stfloats}
\usepackage{url}
\usepackage{verbatim}
\usepackage{graphicx}
\usepackage{microtype}
\usepackage{needspace}
\usepackage{cite}
\newtheorem{definition}{\bf Definition}[section]

\newtheorem{remark}{\bf Remark}[section]

\newtheorem{proposition}{\bf Proposition}[section]

\usepackage{booktabs}
\usepackage{color}
\usepackage{enumerate}
\usepackage{tabularx}
\usepackage{makecell}
\usepackage{multirow}
\usepackage{indentfirst}
\usepackage{epsfig}

\begin{document}
\raggedbottom

\title{Attack-Aware Intrusion Risk Monitoring for Aerial Surveillance via Probabilistic Signal Temporal Logic}

\author{Xinhao~Yan,~\IEEEmembership{Graduate~Student~Member,~IEEE}, 
Kenji~Kashima,~\IEEEmembership{Senior~Member,~IEEE}, 
Bo~Chen,~\IEEEmembership{Senior~Member,~IEEE}, 
Hailong~Huang,~\IEEEmembership{Senior~Member,~IEEE}
\thanks{Xinhao Yan and Hailong Huang are with the Department of Aeronautical and Aviation Engineering, The Hong Kong Polytechnic University, Kowloon, Hong Kong (emails: xin-hao-shawn.yan@connect.polyu.hk; hailong.huang@polyu.edu.hk).}
\thanks{Kenji Kashima is with the Graduate School of Informatics, Kyoto University, Kyoto 606-8501, Japan (email: kk@i.kyoto-u.ac.jp).}
\thanks{Bo Chen is with the Department of Automation, Zhejiang University of Technology, Hangzhou 310023, China (email: bchen@aliyun.com).
}
\thanks{\textit{(Corresponding author: Hailong Huang.)}}
}

\date{}
\markboth{}%
{}

\maketitle

\begin{abstract}
This paper considers online intrusion-risk assessment and graded warning for unmanned aerial vehicles under bounded measurement attacks.
To address this problem, we develop a monitoring framework that combines trusted state estimation, attack-aware risk assessment, and probabilistic signal temporal logic (PrSTL).
Trusted observations support recursive state estimation, while potentially attacked observations inform integrity testing and intrusion-risk assessment.
The key contribution is a confidence-set construction that combines normalized innovation statistics with physical attack bounds to characterize admissible measurement biases.
The resulting family of bias-corrected Gaussian posteriors yields an upper bound on the probability of occupying a no-fly zone or its surrounding safety margin.
Confidence guarantees and finite-cover numerical certificates quantify the reliability of this risk envelope, which is evaluated with or without an integrity alarm.
The PrSTL monitor analyzes the resulting risk sequence to distinguish ordered, persistent, and repeated risk and produces graded warnings with explicit robustness margins and supporting windows.
We validate the monitor in flight experiments with a Crazyflie quadrotor tracked by a Vicon motion capture system.
Controlled comparisons quantify the contributions of individual monitoring components.
Prescribed risk profiles verify the temporal semantics and illustrate how distinct risk patterns produce different warning levels.
\end{abstract}


\section{Introduction}

\subsection{Background}

Unmanned aerial vehicles (UAVs) rely on state estimation~\cite{XinhaoYan_HE_TIV} for a range of missions, including emergency communications and localization~\cite{Tian_EmergencyUAV}, inspection of photovoltaic plants~\cite{Calvo_AerialInspection}, cooperative tracking of moving targets~\cite{Liu_UAVTargetTracking}, and navigation in hazardous low-altitude environments~\cite{Wen_Dangerous}.
At the same time, the proliferation of non-cooperative UAVs operating without prior authorization generates growing risks to public safety, data privacy, and critical infrastructure~\cite{XinhaoYan_Privacy_TASE}.
Such targets may intentionally or unintentionally approach restricted airspace, motivating reliable trajectory estimation, measurement integrity assessment, and timely risk evaluation with respect to prescribed no-fly zones~\cite{Song_Non}.
Liu et al.~\cite{Liu_SequentialThreat_TAES} used sequences of incomplete and noisy UAV observations to assess threat severity and response urgency.
Therefore, effective monitoring must account for residual measurement-attack uncertainty and distinguish transient, sustained, and repeated risk along a trajectory.

\subsection{Related Work}

Distributed estimation methods protect shared state information through confidentiality-preserving fusion~\cite{XinhaoYan_Estimation_TAES} and differential privacy~\cite{XinhaoYan_DP_TAES}.
Zhu et al.~\cite{Zhu_IMMUKF} combined an unscented variant of the Kalman filter (KF) with generative adversarial networks and reinforcement learning to predict unmanned surface vessel trajectories in cooperative aerial--surface operations.
Abdelkader et al.~\cite{Abdelkader_KF} used a KF to guide sensor fusion for robust UAV tracking in dynamic environments.
Pardhasaradhi and Cenkeramaddi~\cite{Pardhasaradhi_KF} integrated distributed radar tracking and fusion with detection and mitigation of Global Positioning System spoofing.
For challenging sea--air backgrounds, Ye et al.~\cite{Ye_KF} adopted a Cauchy-kernel-based adaptive extended KF for UAV target tracking.
Pich\'{e}~\cite{Piche_NIS} examined the normalized innovation squared (NIS) test and a posterior-predictive variant for online assessment of KF consistency.
Or~\cite{Or_GatedNIS_TAES} showed that the covariance of innovations accepted by a validation gate is lower than the nominal innovation covariance even under a correctly specified linear Gaussian model.
Yan et al.~\cite{Yan_STLAware_IMM} developed an interacting multiple-model KF for centralized fusion of measurements from base stations, radars, and cameras operating at different sampling rates.

Measurement attacks, including spoofing and sensor manipulation, threaten the sensing integrity of networked aerial surveillance.
Meng et al.~\cite{Meng_Spoofing} combined model-based anomaly detection with state recovery for acoustic-resonance and electromagnetic attacks on inertial sensors.
Zhang et al.~\cite{Zhang_SensorManipulation} analyzed how sensor-induced errors propagate across interconnected modules and identified overlooked attack vectors.
Bai et al.~\cite{Bai_GNSS5G_TAES} used a single fifth-generation cellular base station to aid innovation-based detection of global navigation satellite system (GNSS) spoofing and to recover the authentic position.
Venturino et al.~\cite{Venturino_ARNF_TAES} combined antenna-array hypothesis testing with adaptive bias estimation to detect and mitigate coordinated spoofing.
Das and Bhaumik~\cite{Das_AngleOnly_TAES} used variational Bayesian inference to estimate target kinematics and attack statistics jointly from bearing measurements corrupted by additive false data.
Complementary work on navigation integrity provides Bayesian bounds on posterior GNSS integrity risk under measurement faults~\cite{Wang_PosteriorIntegrity_TAES} and Gaussian overbounds for heavy-tailed measurement errors~\cite{Yan_GaussianOverbound_TAES}.

Signal temporal logic (STL) specifies spatiotemporal constraints over continuous signals for cyber-physical systems.
Bartocci et al.~\cite{Bartocci_STL} surveyed specification-based monitoring, while Fainekos and Pappas~\cite{Fainekos_STL} developed quantitative robustness semantics.
For stochastic systems, Salamati et al.~\cite{Salamati_STL} considered data-driven verification under STL constraints, whereas Lindemann et al.~\cite{Lindemann_STL} characterized the risk of stochastic systems with respect to temporal-logic specifications.
STL-based control has also been studied with conformal predictions of uncontrollable agents~\cite{Yu_STL} and online synthesis for uncertain systems~\cite{Yu_STL2}.
Optimization-based model checking and informative online monitoring for complex STL formulas were further studied in~\cite{Sato_STL,Zhang_STL}.
Probabilistic signal temporal logic (PrSTL), introduced by Sadigh and Kapoor~\cite{Sadigh_PrSTL}, replaces deterministic atomic predicates with probability constraints and thereby enables risk-aware reasoning under uncertainty.
More recently, Yan et al.~\cite{Yan_STLAware_IMM} combined STL-aware adaptive multiple-model filtering with probabilistic reachable sets for predictive warning of non-cooperative UAVs under uncertain motion modes.

\subsection{Contributions}

The preceding studies leave a specific gap between measurement integrity and temporal intrusion-risk monitoring.
A consistency test or a corrected state estimate alone does not bound regional occupancy risk when an observation retains an undetected bias.
Posterior integrity bounds quantify navigation-error events, while temporal-logic monitoring requires a risk sequence that accounts for spatial constraints and sensing uncertainty.
Therefore, aerial surveillance must incorporate residual bias hypotheses into no-fly-zone occupancy assessments on both detector branches and generate interpretable warnings.
This paper establishes that connection under calibrated Gaussian estimation and physically bounded measurement biases, without assigning an attack prior.
The main contributions are as follows.
\begin{itemize}
\item
We construct a confidence set for unknown measurement biases by combining calibrated innovation statistics with physical attack bounds.
The construction provides pointwise conditional coverage and retains residual bias uncertainty for intrusion-risk assessment even when the integrity detector raises no alarm.
\item
We develop an intrusion-risk envelope by maximizing regional occupancy probabilities over the Gaussian posteriors associated with the retained bias hypotheses.
Confidence analysis and finite-cover numerical certification quantify the reliability of the resulting bounds, while physical attack constraints restore observation dependence and can reduce conservatism.
\item
We establish a PrSTL monitoring mechanism that distinguishes ordered, persistent, and repeated risk.
Boolean satisfaction is linked to positive quantitative robustness, and each graded warning identifies its severity, robustness margin, dominant pattern, and supporting window.
\end{itemize}


\section{System Model and Problem Formulation}

\subsection{State-Space Model}

A non-cooperative aerial target is monitored near restricted regions at intervals of $\Delta t$ over $\mathbb K=\mathbb Z_{\geq0}$.
Its motion and observations are described by the discrete-time model
\begin{equation}
\begin{aligned}
\mathbf x_{k+1}&=\mathbf A\mathbf x_k+\mathbf w_k,\\
\mathbf y_k^0&=\mathbf C\mathbf x_k+\mathbf v_k,
\ \mathbf y_k^a=\mathbf y_k^0+\mathbf a_k^d,\\
\mathbf z_k&=\mathbf H\mathbf x_k+\boldsymbol\eta_k,\ \text{if }\gamma_k=1.
\end{aligned}
\label{eq:state_dynamics}
\end{equation}
The state $\mathbf x_k\in\mathbb R^{n_x}$ contains planar position $\mathbf p_k\in\mathbb R^2$, velocity $\dot{\mathbf p}_k\in\mathbb R^2$, and optional kinematic states.
The vulnerable observation $\mathbf y_k^a\in\mathbb R^{n_y}$ is available at every step, while the protected but noisy observation $\mathbf z_k\in\mathbb R^{n_z}$ is available only when the prescribed indicator $\gamma_k=1$.
Here, we use the deterministic periodic schedule $\gamma_k=\mathbf1\{k\bmod m=0\}$ with integer $m\geq1$, and the two sampling intervals are $\Delta t$ and $m\Delta t$.
At a scheduled step, $\mathbf z_k$ is assimilated after the risk assessment to support the next recursive estimation step.

The process noise $\mathbf w_k$ and the measurement noises $\mathbf v_k$ and $\boldsymbol\eta_k$ are zero-mean Gaussian with covariances $\mathbf Q\succeq0$, $\mathbf R\succ0$, and $\mathbf R^t\succ0$, respectively.
These noises are mutually independent, independent across time, and independent of the initial state.
The initial state follows $\mathbf x_0\sim\mathcal N(\hat{\mathbf x}_{0|-1},\mathbf P_{0|-1})$, where $\mathcal N(\boldsymbol\mu,\mathbf P)$ denotes a Gaussian distribution with mean $\boldsymbol\mu$ and covariance $\mathbf P$.
Prediction covariances are assumed positive definite so that the corresponding Gaussian densities exist.
The recursive estimator uses the available trusted observations, represented by the information set
\begin{equation}
\mathcal I_{k-1}:=\sigma_{\rm alg}(\mathbf z_j:0\leq j<k,\ \gamma_j=1),
\label{eq:trusted_information}
\end{equation}
with trivial initial information $\mathcal I_{-1}$.
Here $\sigma_{\rm alg}(\cdot)$ denotes the sigma-algebra generated by the listed observations.

\begin{figure}[t]
\centering
\includegraphics[width=1.0\columnwidth]{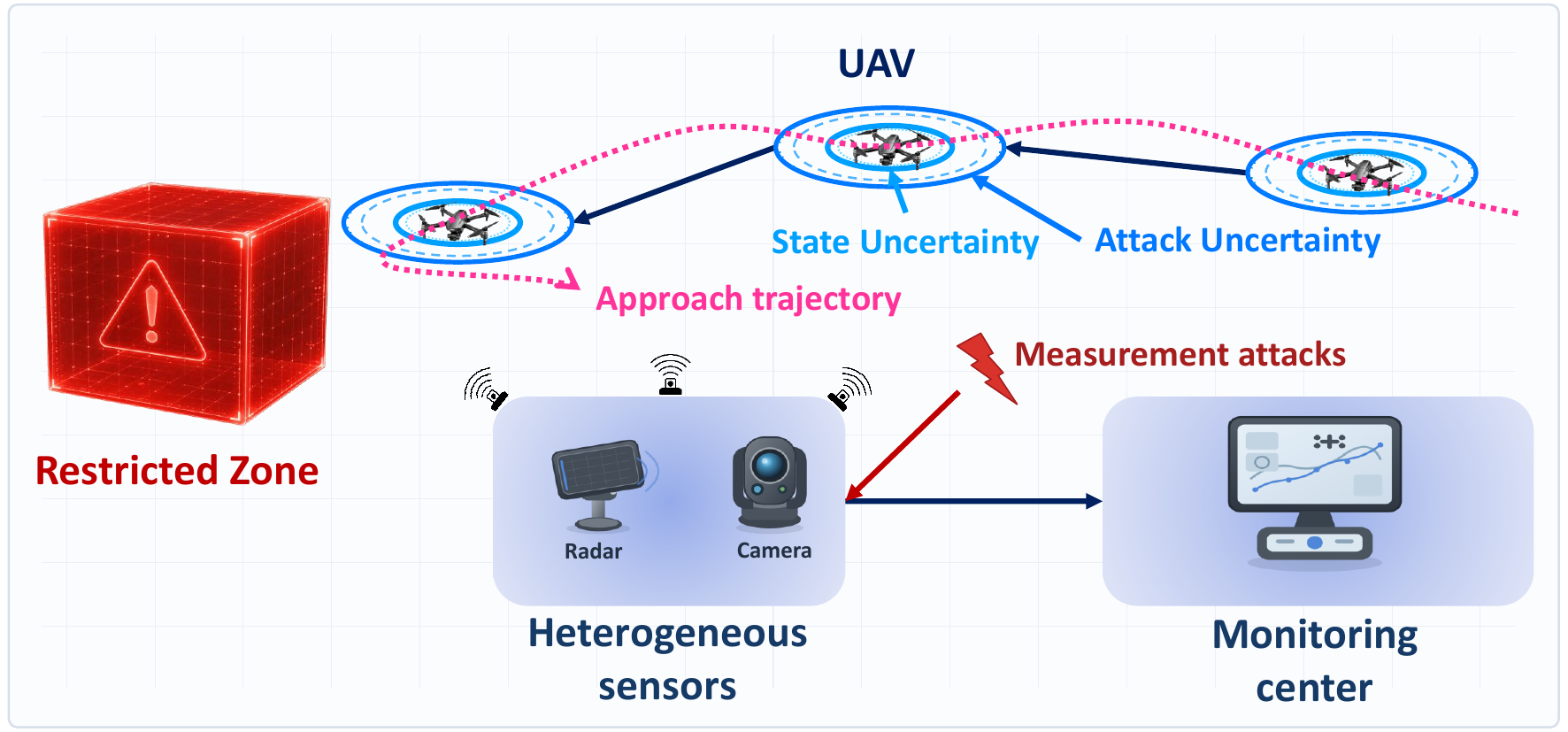}%
\caption{UAV intrusion-risk monitoring with heterogeneous sensing subject to measurement attacks.
State uncertainty and attack uncertainty are considered when assessing entry into a restricted zone.}
\label{Fig_Structure}
\end{figure}

The unknown bias $\mathbf a_k^d$ may depend on $\mathcal I_{k-1}$ and public trusted priors, but not on the current private state, current noise, or any additional history of vulnerable observations.
Therefore, the bias is treated as an unknown $\mathcal I_{k-1}$-measurable quantity without an assigned prior distribution.
Available physical knowledge specifies a known $\mathcal I_{k-1}$-measurable, nonempty compact convex set $\mathcal A_k\subset\mathbb R^{n_y}$ before $\mathbf y_k^a$ is observed, with $\mathbf0,\mathbf a_k^d\in\mathcal A_k$.
This constraint is justified independently of the current observation and may take the form $\mathcal A_k=\{\mathbf a:\|\mathbf a\|_2\leq a_{\max}\}$ for a physical bias bound $a_{\max}$.
Fig.~\ref{Fig_Structure} illustrates UAV intrusion-risk monitoring under state uncertainty and measurement attacks.

\subsection{Trusted State Estimation and Monitoring Objective}

Starting from the initial Gaussian, the trusted posterior is propagated for $k\geq1$ according to
\begin{equation}
\begin{aligned}
\hat{\mathbf x}_{k|k-1}&=\mathbf A\hat{\mathbf x}_{k-1|k-1}^{t},\\
\mathbf P_{k|k-1}&=\mathbf A\mathbf P_{k-1|k-1}^{t}\mathbf A^\top+\mathbf Q.
\end{aligned}
\label{eq:kf_prediction}
\end{equation}
When $\gamma_k=1$, the innovation covariance is defined as $\mathbf S_k^t=\mathbf H\mathbf P_{k|k-1}\mathbf H^\top+\mathbf R^t$, and the trusted correction is performed after risk assessment using
\begin{equation}
\begin{aligned}
\mathbf K_k^t&=\mathbf P_{k|k-1}\mathbf H^\top(\mathbf S_k^t)^{-1},\\
\hat{\mathbf x}_{k|k}^{t}&=\hat{\mathbf x}_{k|k-1}
 +\mathbf K_k^t(\mathbf z_k-\mathbf H\hat{\mathbf x}_{k|k-1}),\\
\mathbf P_{k|k}^{t}&=\mathbf P_{k|k-1}-\mathbf K_k^t\mathbf S_k^t(\mathbf K_k^t)^\top.
\end{aligned}
\label{eq:trusted_kf_update}
\end{equation}
When $\gamma_k=0$, take $\hat{\mathbf x}_{k|k}^{t}=\hat{\mathbf x}_{k|k-1}$ and $\mathbf P_{k|k}^{t}=\mathbf P_{k|k-1}$.

Gaussian propagation and conditioning on the available trusted observations give the conditional density
\begin{equation}
p(\mathbf x_k\mid\mathcal I_{k-1})
=\mathcal N(\mathbf x_k;\hat{\mathbf x}_{k|k-1},\mathbf P_{k|k-1}).
\label{eq:trusted_calibration}
\end{equation}
Here, $p(\cdot\mid\mathcal I_{k-1})$ denotes a conditional density, and $\mathcal N(\mathbf x;\boldsymbol\mu,\mathbf P)$ is the Gaussian density evaluated at $\mathbf x$.
The identity holds inductively for both scheduled corrections and prediction-only steps.
The prediction error $\mathbf e_{k|k-1}=\mathbf x_k-\hat{\mathbf x}_{k|k-1}$ is conditionally independent of $\mathbf v_k$, because the recursion uses only process and trusted-sensor information.

The current vulnerable observation is assessed using the innovation, innovation covariance, and Kalman gain, which are defined by
\begin{equation}
\begin{aligned}
\boldsymbol\epsilon_k^a&=\mathbf y_k^a-\mathbf C\hat{\mathbf x}_{k|k-1},\\
\mathbf S_k&=\mathbf C\mathbf P_{k|k-1}\mathbf C^\top+\mathbf R,\\
\mathbf K_k&=\mathbf P_{k|k-1}\mathbf C^\top\mathbf S_k^{-1}.
\end{aligned}
\label{eq:kf_innovation}
\end{equation}
The nominal innovation $\boldsymbol\epsilon_k^0=\mathbf C\mathbf e_{k|k-1}+\mathbf v_k$ has conditional density $p(\boldsymbol\epsilon_k^0\mid\mathcal I_{k-1})=\mathcal N(\boldsymbol\epsilon_k^0;\mathbf0,\mathbf S_k)$.
Therefore, the NIS value $b_k^a=(\boldsymbol\epsilon_k^a)^\top\mathbf S_k^{-1}\boldsymbol\epsilon_k^a$ defines the integrity detector
\begin{equation}
h_\alpha=F_{\chi_{n_y}^2}^{-1}(1-\alpha),\
\mathcal D_k=\mathbf1\{b_k^a>h_\alpha\}.
\label{eq:chi_detector}
\end{equation}
If there is no attack, the detector has conditional false-alarm probability $\alpha\in(0,1)$.
For a fixed current bias $\mathbf a$, its repeated-sampling no-alarm probability is $q_k^{\rm nd}(\mathbf a)=F_{\chi_{n_y}^{\prime2}(\lambda_k(\mathbf a))}(h_\alpha)$, with noncentrality parameter $\lambda_k(\mathbf a)=\mathbf a^\top\mathbf S_k^{-1}\mathbf a$.

Conditioning on the vulnerable observation for a fixed bias hypothesis yields the covariance
\begin{equation}
\mathbf P_{k|k}^{+}=\mathbf P_{k|k-1}-\mathbf K_k\mathbf S_k\mathbf K_k^\top.
\label{eq:kf_measurement_update}
\end{equation}
This covariance is used for risk assessment, while only trusted observations enter the recursive estimator.
The monitoring objective is to bound occupancy of a no-fly zone or its surrounding safety margin using the available observation and bias information.
Since a no-alarm decision does not resolve the decomposition $\boldsymbol\epsilon_k^a=\boldsymbol\epsilon_k^0+\mathbf a_k^d$, the assessment must retain plausible biases on both detector branches.

\section{Detector-Consistent Risk Assessment and Temporal Monitoring}

\subsection{Attack Confidence Set}

For a bias hypothesis $\mathbf a$, the corrected innovation is defined as $\boldsymbol\epsilon_k(\mathbf a)=\boldsymbol\epsilon_k^a-\mathbf a$.
Inverting the calibrated chi-square pivot at miscoverage level $\delta\in(0,1)$, with $h_\delta=F_{\chi^2_{n_y}}^{-1}(1-\delta)$, produces the confidence ellipsoid and physically constrained set
\begin{equation}
\begin{aligned}
\mathcal E_k^a(\delta)
&=\{\mathbf a:(\boldsymbol\epsilon_k^a-\mathbf a)^\top\mathbf S_k^{-1}
(\boldsymbol\epsilon_k^a-\mathbf a)\leq h_\delta\},\\
\mathcal C_k^a(\delta)&=\mathcal E_k^a(\delta)\cap\mathcal A_k.
\end{aligned}
\label{eq:constrained_attack_set}
\end{equation}
The ellipsoid is compact and convex, and its intersection with $\mathcal A_k$ may be empty.
At the realized bias, $\boldsymbol\epsilon_k(\mathbf a_k^d)=\boldsymbol\epsilon_k^0$.
The calibration in \eqref{eq:trusted_calibration}, together with $\mathbf a_k^d\in\mathcal A_k$, establishes exact pointwise coverage:
\begin{equation}
\mathbb P\{\mathbf a_k^d\in\mathcal C_k^a(\delta)\mid\mathcal I_{k-1}\}=1-\delta.
\label{eq:attack_coverage}
\end{equation}
The probability in \eqref{eq:attack_coverage} concerns repeated observations conditional on the available trusted information.
The observed set itself does not define an attack posterior.
Further conditioning on no alarm selects a subset of observations, for which $q_k^{\rm nd}=q_k^{\rm nd}(\mathbf a_k^d)>0$ gives the coverage bound
\begin{equation}
\begin{aligned}
&\mathbb P\{\mathbf a_k^d\in\mathcal C_k^a(\delta)\mid\mathcal I_{k-1},\mathcal D_k=0\}\\
&\geq\max\{0,1-\delta/q_k^{\rm nd}\}.
\end{aligned}
\label{eq:no_alarm_selected_coverage}
\end{equation}
If $\mathcal D_k=0$ and $\delta\leq\alpha$, then $b_k^a\leq h_\alpha\leq h_\delta$ and $\mathbf0\in\mathcal A_k$ imply $\mathbf0\in\mathcal C_k^a(\delta)$.
Thus, zero bias remains feasible after no alarm, while $\alpha$ and $\delta$ separately control false alarms and confidence-set miscoverage.

For the subsequent risk maximization, the same constraints can be expressed in terms of corrected residuals:
\begin{equation}
\begin{aligned}
\mathcal R_k(\delta)&=\{\mathbf r:\mathbf r^\top\mathbf S_k^{-1}\mathbf r\leq h_\delta\},\\
\mathcal R_k^{\mathcal A}(\delta)&=\{\mathbf r\in\mathcal R_k(\delta):
\boldsymbol\epsilon_k^a-\mathbf r\in\mathcal A_k\}.
\end{aligned}
\label{eq:constrained_residual_set}
\end{equation}
The bijection $\mathbf r=\boldsymbol\epsilon_k^a-\mathbf a$ maps $\mathcal C_k^a(\delta)$ onto $\mathcal R_k^{\mathcal A}(\delta)$.

\subsection{Attack-Aware Intrusion Risk}

A bias-corrected vulnerable observation defines the candidate posterior as
\begin{equation}
\begin{aligned}
\hat{\mathbf x}_{k|k}(\mathbf a)
&=\hat{\mathbf x}_{k|k-1}+\mathbf K_k(\boldsymbol\epsilon_k^a-\mathbf a),\\
\mathbf X_k^{\mathbf a}&\sim\mathcal N(\hat{\mathbf x}_{k|k}(\mathbf a),\mathbf P_{k|k}^{+}).
\end{aligned}
\label{eq:attack_conditioned_posterior}
\end{equation}
Since the realized bias is fixed given $\mathcal I_{k-1}$, Gaussian conditioning yields $p(\mathbf x_k\mid\mathcal I_{k-1},\mathbf y_k^a)=\mathcal N(\mathbf x_k;\hat{\mathbf x}_{k|k}(\mathbf a_k^d),\mathbf P_{k|k}^{+})$.
Collecting the retained hypotheses gives the following common-covariance posterior family:
\begin{equation}
\mathcal X_k=\{\mathcal N(\hat{\mathbf x}_{k|k}(\mathbf a),\mathbf P_{k|k}^{+}):
\mathbf a\in\mathcal C_k^a(\delta)\}.
\label{eq:posterior_family}
\end{equation}

Let $\mathcal Z\subset\mathbb R^2$ be a bounded restricted region with nonempty interior and piecewise-smooth boundary, represented by a simple polygon, circle, or ellipse.
Write $d(\mathbf p,\mathcal Z)$ for the distance to $\operatorname{cl}(\mathcal Z)$ outside the region and minus the distance to $\operatorname{cl}(\mathcal Z^c)$ inside it.
For safety distance $th^d\geq0$, define $\theta_{\mathcal Z}(\mathbf p)=d(\mathbf p,\mathcal Z)-th^d$ and the associated safety region $\mathcal Z_{th^d}=\{\mathbf p:\theta_{\mathcal Z}(\mathbf p)<0\}$.
The matrix $\mathbf C_{\rm pos}$ extracts planar position, $\mathbf p_k=\mathbf C_{\rm pos}\mathbf x_k$.
On the realized state trace $\mathbf x=(\mathbf x_k)_{k\in\mathbb K}$, the safety atom $\mu_{\mathcal Z}^{\rm safe}$ holds when $\theta_{\mathcal Z}(\mathbf p_k)\geq0$.
For $I=[a,b]$, $\psi_{\mathcal Z}=\mathcal G_I\mu_{\mathcal Z}^{\rm safe}$, and indices $\mathbb K_k(I)=[k+a,k+b]\cap\mathbb K$, standard STL quantifies the spatial safety margin by
\begin{equation}
\begin{aligned}
\rho_{\rm det}(\mathbf x,\mu_{\mathcal Z}^{\rm safe},k)
&=\theta_{\mathcal Z}(\mathbf C_{\rm pos}\mathbf x_k),\\
\rho_{\rm det}(\mathbf x,\psi_{\mathcal Z},k)
&=\min_{j\in\mathbb K_k(I)}\theta_{\mathcal Z}(\mathbf C_{\rm pos}\mathbf x_j).
\end{aligned}
\label{eq:realized_stl_robustness}
\end{equation}
Positive, negative, and zero margins indicate clearance, intrusion, and boundary contact, respectively.
Conditional state distributions quantify uncertainty about these realized-state predicates.

The position variable $\mathbf p_k^{\mathbf a}=\mathbf C_{\rm pos}\mathbf X_k^{\mathbf a}$ has Gaussian moments expressed as
\begin{equation*}
\begin{aligned}
\hat{\mathbf p}_k(\mathbf a)&=\mathbf C_{\rm pos}\hat{\mathbf x}_{k|k}(\mathbf a),\\
\boldsymbol\Sigma_k^{\rm pos}&=\mathbf C_{\rm pos}\mathbf P_{k|k}^{+}\mathbf C_{\rm pos}^{\top}.
\end{aligned}
\end{equation*}
The position covariance is assumed positive definite at the evaluated indices.
Integrating the bivariate Gaussian density $\phi_2$ of $\mathbf p_k^{\mathbf a}$ over the safety region gives the intrusion probability
\begin{equation}
P_k^{\rm in}(\mathbf a)
=\int_{\mathcal Z_{th^d}}\phi_2(\mathbf p;\hat{\mathbf p}_k(\mathbf a),\boldsymbol\Sigma_k^{\rm pos})\,d\mathbf p.
\label{eq:attack_conditioned_intrusion_probability}
\end{equation}
More generally, for a Borel-measurable state predicate $\vartheta$, the retained hypotheses induce the upper event probability
\begin{equation}
\overline P_k[\vartheta]=
\begin{cases}
\displaystyle\sup_{\mathbf a\in\mathcal C_k^a(\delta)}
\mathbb P\{\vartheta(\mathbf X_k^{\mathbf a})<0\},&\mathcal C_k^a(\delta)\ne\varnothing,\\
1,&\mathcal C_k^a(\delta)=\varnothing.
\end{cases}
\label{eq:posterior_upper_probability}
\end{equation}
If the feasible set is empty, the monitor returns the conservative value $1$, which does not imply that an intrusion has occurred.
With $\vartheta_{\mathcal Z}(\mathbf x)=\theta_{\mathcal Z}(\mathbf C_{\rm pos}\mathbf x)$, define $P_k^{pr,{\rm intru}}=\overline P_k[\vartheta_{\mathcal Z}]$.
The confidence-set coverage yields the following risk guarantee.

\begin{proposition}[\textbf{Risk Guarantee}]\label{prop:risk_coverage}
Suppose the stated sensing and attack conditions hold, the trusted predictor satisfies \eqref{eq:trusted_calibration}, and $\boldsymbol\Sigma_k^{\rm pos}\succ\mathbf0$.
Then the risk envelope attains its maximum whenever $\mathcal C_k^a(\delta)\ne\varnothing$.
At each index, risk underestimation has conditional probability at most $\delta$ given $\mathcal I_{k-1}$.
For a prescribed finite index set $\mathcal J\subset\mathbb K$, use predetermined levels $\delta_k$ in the confidence-set construction, with $\delta_k=\delta$ for a common level.
The envelopes then satisfy the simultaneous bound
\begin{equation}
\begin{aligned}
&\mathbb P\bigl\{\forall k\in\mathcal J:
P_k^{\rm in}(\mathbf a_k^d)\leq P_k^{pr,{\rm intru}}\bigr\}\\
&\ \geq1-\sum_{k\in\mathcal J}\delta_k.
\end{aligned}
\label{eq:risk_coverage}
\end{equation}
At an individual index, the joint event of $\mathcal D_k=0$ and risk underestimation has conditional probability at most $\delta_k$.
If $\mathcal D_k=0$ and $\delta_k\leq\alpha$, then $P_k^{pr,{\rm intru}}\geq P_k^{\rm in}(\mathbf0)$.
\end{proposition}
\Needspace{3\baselineskip}
\begin{IEEEproof}
Continuity of Gaussian translations in $L^1$ makes $P_k^{\rm in}(\mathbf a)$ continuous on the compact feasible set.
On $\mathbf a_k^d\in\mathcal C_k^a(\delta_k)$, the true bias is included in the maximization, so \eqref{eq:attack_coverage} gives the pointwise conditional bound.
Taking expectations and applying the union bound yields \eqref{eq:risk_coverage} without requiring independence across indices.
The empty-set output is conservative.
Intersecting failure with no alarm gives the joint bound, and retention of $\mathbf0$ proves nominal dominance.
\end{IEEEproof}

\begin{remark}
\label{rem:measurement_invariance}
Without the physical constraint, $\mathbf r=\boldsymbol\epsilon_k^a-\mathbf a$ gives the posterior mean family
$\{\hat{\mathbf x}_{k|k-1}+\mathbf K_k\mathbf r:\mathbf r\in\mathcal R_k(\delta)\}$.
For a fixed trusted prior, this family and its maximal occupancy are observation-invariant.
The physical constraint restores observation dependence through $\boldsymbol\epsilon_k^a-\mathbf r\in\mathcal A_k$, replacing $\mathcal R_k$ by $\mathcal R_k^{\mathcal A}$.
\end{remark}

The monitor applies $P_k^{\rm mon}[\vartheta]=\overline P_k[\vartheta]$ on both detector branches and reports the risk sequence alongside the integrity indicator $I_k^{\rm int}=\mathcal D_k$.
The guarantee concerns the accuracy of the probability envelope rather than the occurrence of trajectory-level intrusion.

For two-dimensional measurements, choose $\mathbf D_k\mathbf D_k^\top=h_\delta\mathbf S_k$ and define $\mathbf G_k=\mathbf C_{\rm pos}\mathbf K_k\mathbf D_k$ and $\boldsymbol\mu_k^{0}=\mathbf C_{\rm pos}\hat{\mathbf x}_{k|k-1}$.
A retained bias can be written as $\mathbf a=\boldsymbol\epsilon_k^a-\mathbf D_k\mathbf u$, where $\|\mathbf u\|_2\leq1$ and $\mathbf a\in\mathcal A_k$.
The corresponding position mean is $\boldsymbol\mu_k^{0}+\mathbf G_k\mathbf u$.
A finite cover of the unit disk uses square cells of half-width $1/M$ and centers $\mathbf c_j$.
The within-cell mean displacement is bounded by $\eta_k=\max_{\mathbf s\in\{-1/M,1/M\}^2}\|\mathbf G_k\mathbf s\|_2$.
The $1$-Lipschitz signed distance and a common-noise Gaussian coupling give the following bound within cell $j$:
\begin{equation}
\begin{aligned}
P_k^{\rm in}(\boldsymbol\epsilon_k^a-\mathbf D_k\mathbf u)
&\leq\mathbb P\{d(\mathbf p_{j,k},\mathcal Z)<th^d+\eta_k\}=:p_{j,k}^+,\\
\mathbf p_{j,k}&\sim\mathcal N(\boldsymbol\mu_k^{0}+\mathbf G_k\mathbf c_j,\boldsymbol\Sigma_k^{\rm pos}).
\end{aligned}
\label{eq:numerical_inflated_probability}
\end{equation}
Cells are removed only when their intersection with the feasible domain is certified empty.
For the norm bound, cells can be excluded whenever
\begin{equation*}
\|\boldsymbol\epsilon_k^a-\mathbf D_k\mathbf c_j\|_2>a_{\max}+\max_{\mathbf s\in\{-1/M,1/M\}^2}\|\mathbf D_k\mathbf s\|_2.
\end{equation*}

Let $s_{j,k}$ count occurrences of the inflated-region event in $N$ independent and identically distributed (i.i.d.) Gaussian draws.
A one-sided Clopper-Pearson upper confidence endpoint for $p_{j,k}^+$ is computed as
\begin{equation}
U_{j,k}=\begin{cases}
F_{{\rm Beta}(s_{j,k}+1,N-s_{j,k})}^{-1}(1-q),&s_{j,k}<N,\\
1,&s_{j,k}=N,
\end{cases}
\label{eq:numerical_cp_upper}
\end{equation}
where $q=\beta/N_{\rm tests}$ and $N_{\rm tests}$ is a predetermined upper bound on the number of combinations of cells, regions, and indices to be evaluated.
This bound remains fixed after pruning.
A deterministic bound from the standard bivariate Gaussian radial tail is given by
\begin{equation*}
p_{j,k}^+\leq\exp\!\left[-\frac{d_{j,k}^2}{2\lambda_{\max}(\boldsymbol\Sigma_k^{\rm pos})}\right],
\end{equation*}
where $d_{j,k}$ is the distance from the mean at the center of cell $j$ to the bounding box of the inflated region.
This bound can tighten the endpoint.
Taking the maximum of the resulting upper endpoint and a feasible-witness estimate preserves validity.
Hereafter, $U_{j,k}$ denotes the resulting certified endpoint.

The union bound ensures $\widehat P_k^{pr,{\rm intru}}=\max_jU_{j,k}\geq P_k^{pr,{\rm intru}}$ simultaneously over the preallocated evaluations with nonempty retained sets, with probability at least $1-\beta$.
The monitor returns $1$ if the feasible set is certified empty, whereas coverage of the true bias ensures that the set is nonempty.
The i.i.d. draws are independent of system, sensing, and attack data.
Reusing them across cells, regions, indices, or paired calls preserves the union bound.
Combining numerical and confidence-set failure events over the prescribed index set $\mathcal J$ establishes the total error bound
\begin{equation}
\begin{aligned}
&\mathbb P\bigl\{\exists k\in\mathcal J:
P_k^{\rm in}(\mathbf a_k^d)>\widehat P_k^{pr,{\rm intru}}\bigr\}\\
&\ \leq\sum_{k\in\mathcal J}\delta_k+\beta.
\end{aligned}
\label{eq:numerical_total_failure}
\end{equation}
Here $\beta$ is shared over all preallocated evaluations in $\mathcal J$.
For a single index, the corresponding conditional failure bound is $\delta_k+\beta$ given $\mathcal I_{k-1}$.
The certified risk sequence supplies the numerical input to the PrSTL monitor.

\subsection{PrSTL Syntax and Semantics}

Let $\mathcal X=(\mathcal X_j)_{j=0}^{H}$ denote the sequence of posterior families defined by \eqref{eq:posterior_family}, where $H$ is the last sample index.
In PrSTL, probability constraints concern marginal state events, while temporal operators express their ordering and duration along the sequence.
The syntax is defined below.

\begin{definition}[\textbf{PrSTL Syntax}]\label{def:prstl_syntax}
For $I=[a,b]$ with integer endpoints $0\leq a\leq b$, formulas are generated by
\begin{equation}
\begin{aligned}
\varphi::={}&\top\mid\mu_{\vartheta,\epsilon}^{pr}
\mid\tilde\neg\mu_{\vartheta,\epsilon}^{pr}
\mid\varphi_1\land\varphi_2\mid\varphi_1\lor\varphi_2\\
&\mid\varphi_1\mathcal U_I\varphi_2
\mid\mathcal F_I\varphi\mid\mathcal G_I\varphi.
\end{aligned}
\label{eq:STL_Syntax}
\end{equation}
Here $\vartheta:\mathbb R^{n_x}\to\mathbb R$ is Borel-measurable, $\epsilon\in[0,1]$, and $\mathcal U$, $\mathcal F$, and $\mathcal G$ denote until, eventually, and always.
\end{definition}
The required interval extent $h(\varphi)$ is zero for atoms and $\top$, takes the maximum for Boolean combinations, and obeys
\begin{equation*}
\begin{aligned}
h(\mathcal O_{[a,b]}\varphi)&=b+h(\varphi),\ \mathcal O\in\{\mathcal F,\mathcal G\},\\
h(\varphi_1\mathcal U_{[a,b]}\varphi_2)&=b+\max\{h(\varphi_1),h(\varphi_2)\}.
\end{aligned}
\end{equation*}
Complete evaluation requires $j+h(\varphi)\leq H$, with temporal indices
\begin{equation}
\mathbb K_j(I)=[j+a,j+b]\cap\mathbb K,\ 
\mathbb J(j,u)=[j,u)\cap\mathbb Z.
\label{eq:temporal_index_sets}
\end{equation}
These indices determine the Boolean and quantitative semantics below.

\begin{definition}[\textbf{PrSTL Semantics}]\label{def:prstl_semantics}
The formula $\top$ is always true, and conjunction and disjunction follow their usual Boolean rules.
At index $j$, $\mu_{\vartheta,\epsilon}^{pr}$ holds iff $P_j^{\rm mon}[\vartheta]>1-\epsilon$, and its predicate reversal holds iff $P_j^{\rm mon}[-\vartheta]>1-\epsilon$.
The formula $\mathcal F_I\varphi$ ($\mathcal G_I\varphi$) holds iff $\varphi$ holds at some (every) $u\in\mathbb K_j(I)$.
The formula $\varphi_1\mathcal U_I\varphi_2$ holds iff some $u\in\mathbb K_j(I)$ satisfies $\varphi_2$ and $\varphi_1$ holds throughout $\mathbb J(j,u)$.

Write $\rho_\varphi(j)=\rho(\mathcal X,\varphi,j)\in[-1,1]$ for the probability margin, distinct from the spatial robustness in \eqref{eq:realized_stl_robustness}.
Set $\rho_\top(j)=1$, with the atomic and Boolean margins defined by
\begin{equation}
\begin{aligned}
\rho_{\mu_{\vartheta,\epsilon}^{pr}}(j)&=P_j^{\rm mon}[\vartheta]-(1-\epsilon),\\
\rho_{\tilde\neg\mu_{\vartheta,\epsilon}^{pr}}(j)&=P_j^{\rm mon}[-\vartheta]-(1-\epsilon),\\
\rho_{\varphi_1\land\varphi_2}(j)&=\min\{\rho_{\varphi_1}(j),\rho_{\varphi_2}(j)\},\\
\rho_{\varphi_1\lor\varphi_2}(j)&=\max\{\rho_{\varphi_1}(j),\rho_{\varphi_2}(j)\}.
\end{aligned}
\label{eq:quantitative_atomic_boolean}
\end{equation}
The temporal margins are expressed as
\begin{equation}
\begin{aligned}
\rho_{\mathcal F_I\varphi}(j)&=\max_{u\in\mathbb K_j(I)}\rho_\varphi(u),\ 
\rho_{\mathcal G_I\varphi}(j)=\min_{u\in\mathbb K_j(I)}\rho_\varphi(u),\\
\rho_{\varphi_1\mathcal U_I\varphi_2}(j)
&=\max_{u\in\mathbb K_j(I)}\min\!\left\{\rho_{\varphi_2}(u),
\min_{v\in \mathbb J(j,u)}\rho_{\varphi_1}(v)\right\},
\end{aligned}
\label{eq:quantitative_temporal_operators}
\end{equation}
where $\min\varnothing=1$ for an empty intervening interval.
\end{definition}

Different maximizing hypotheses and possible probability mass at $\vartheta=0$ prevent predicate reversal from being logical complementation.
The original atom and its reversal may both hold.
Strict inequalities and complete evaluation windows give the following relation between Boolean satisfaction and the sign of the robustness margin.
\begin{proposition}[\textbf{Sign Consistency}]\label{prop:boolean_quantitative_consistency}
For every formula with $j+h(\varphi)\leq H$,
\begin{equation}
(\mathcal X,j)\models\varphi\ \Longleftrightarrow\ 
\rho(\mathcal X,\varphi,j)>0.
\label{eq:boolean_quantitative_consistency}
\end{equation}
\end{proposition}
\Needspace{3\baselineskip}
\begin{IEEEproof}
Structural induction applies because finite minima are positive iff all arguments are positive, and maxima iff some argument is positive, including the empty-interval convention.
\end{IEEEproof}

\subsection{Temporal Warning Mechanism}

For each region, let $s=(s(j))_{j=0}^{H}$ be the risk sequence with $s(j)=P_j^{\rm mon}[\vartheta_{\mathcal Z}]$ or its numerically certified upper bound $\widehat P_j^{pr,{\rm intru}}$, with the guarantee in \eqref{eq:numerical_total_failure}.
The low, medium, and high risk thresholds $0<\tau^L<\tau^M<\tau^H<1$ define the scalar atoms and margins
\begin{equation}
\begin{aligned}
\mu^L &: s(j)<\tau^L,&\rho(s,\mu^L,j)&=\tau^L-s(j),\\
\mu^M &: s(j)>\tau^M,&\rho(s,\mu^M,j)&=s(j)-\tau^M,\\
\mu^H &: s(j)>\tau^H,&\rho(s,\mu^H,j)&=s(j)-\tau^H.
\end{aligned}
\label{eq:warning_atomic_robustness}
\end{equation}
The intermediate band $\mu^I:\tau^L\leq s(j)\leq\tau^M$ activates none of these atoms.
The ordered patterns in $\mathcal Q$ are low then medium ($LM$), medium then high ($MH$), and low then medium then high ($LMH$).
The remaining patterns represent sustained medium ($M$) or high ($H$) risk and repeated medium ($RM$) or high ($RH$) risk at distinct offsets.
For each $q\in\mathcal Q$, the following formulas use window lengths from a nonempty finite set $\mathcal T^q\subset\mathbb Z_{>0}$:
\begin{equation}
\begin{aligned}
\Phi_T^{LM}&=\mathcal F_{[0,T]}
(\mu^L\land\mathcal F_{[1,T]}\mu^M),\\
\Phi_T^M&=\mathcal G_{[0,T]}\mu^M,\ \Phi_T^H=\mathcal G_{[0,T]}\mu^H,\\
\Phi_T^{RM}&=\mathcal F_{[0,T]}
(\mu^M\land\mathcal F_{[1,T]}\mu^M),\\
\Phi_T^{MH}&=\mathcal F_{[0,T]}
(\mu^M\land\mathcal F_{[1,T]}\mu^H),\\
\Phi_T^{LMH}&=\mathcal F_{[0,T]}\big(\mu^L\land
\mathcal F_{[1,T]}\\
&\ (\mu^M\land\mathcal F_{[1,T]}\mu^H)\big),\\
\Phi_T^{RH}&=\mathcal F_{[0,T]}
(\mu^H\land\mathcal F_{[1,T]}\mu^H).
\end{aligned}
\label{eq:warning_patterns}
\end{equation}
Since $\mu^H\Rightarrow\mu^M$, ordering need not imply strict increase, and repetition may overlap persistence.
The required interval extent is $3T$ for $LMH$, $T$ for $M,H$, and $2T$ otherwise, with $T+1$ samples in $[0,T]$.

At an evaluation index $k$ with $k+h(\Phi_T^q)\leq H$ for all configured patterns and windows, the pattern margins and truth values are given by
\begin{equation}
\begin{aligned}
\rho_{\rm pat}(s,q,k)&=\max_{T\in\mathcal T^q}\rho(s,\Phi_T^q,k),\\
\chi_k^q&\Longleftrightarrow\rho_{\rm pat}(s,q,k)>0.
\end{aligned}
\label{eq:pattern_quantitative_robustness}
\end{equation}
Assign severities $\ell(LM)=1$, $\ell(M)=\ell(RM)=2$, $\ell(MH)=\ell(LMH)=3$, and $\ell(H)=\ell(RH)=4$, independently of $\alpha,\delta,\beta$.
Select the highest active severity and then its largest margin:
\begin{equation}
\begin{aligned}
W_k&=\max\big(\{0\}\cup\{\ell(q):q\in\mathcal Q,\ \chi_k^q\}\big),\\
\mathcal Q_k^W&=\begin{cases}\mathcal Q,&W_k=0,\\
\{q\in\mathcal Q:\ell(q)=W_k\},&W_k>0,\end{cases}\\
q_k^\star&\in\arg\max_{q\in\mathcal Q_k^W}\rho_{\rm pat}(s,q,k),\\
T_k^\star&\in\arg\max_{T\in\mathcal T^{q_k^\star}}\rho(s,\Phi_T^{q_k^\star},k),\\
\rho_{\rm warn}(s,k)&=\rho_{\rm pat}(s,q_k^\star,k).
\end{aligned}
\label{eq:quantitative_warning_output}
\end{equation}
Thus, $W_k>0$ iff $\rho_{\rm warn}(s,k)>0$, with the largest nonpositive margin reported otherwise.
Ties are resolved first by pattern in the order $LM,M,RM,MH,LMH,H,RH$, then by increasing $T$, and finally by region index.
The output $\mathcal O_k^{\rm warn}=(W_k,\rho_{\rm warn}(s,k),q_k^\star,T_k^\star)$ reports severity, margin, pattern, and window, with maximum severity displayed across regions.

\begin{remark}
\label{rem:kalman_prediction}
For prediction, propagate each retained Gaussian by $(\mathbf m,\mathbf P)\mapsto(\mathbf A\mathbf m,\mathbf A\mathbf P\mathbf A^\top+\mathbf Q)$ and apply the same risk construction and temporal semantics.
One confidence-set event bounds all marginal risks propagated from a common posterior family.
\end{remark}

\section{Experimental Results}

\subsection{Experimental Setup and Evaluation Protocol}

\begin{figure}[t]
\centering
\includegraphics[width=1.0\columnwidth]{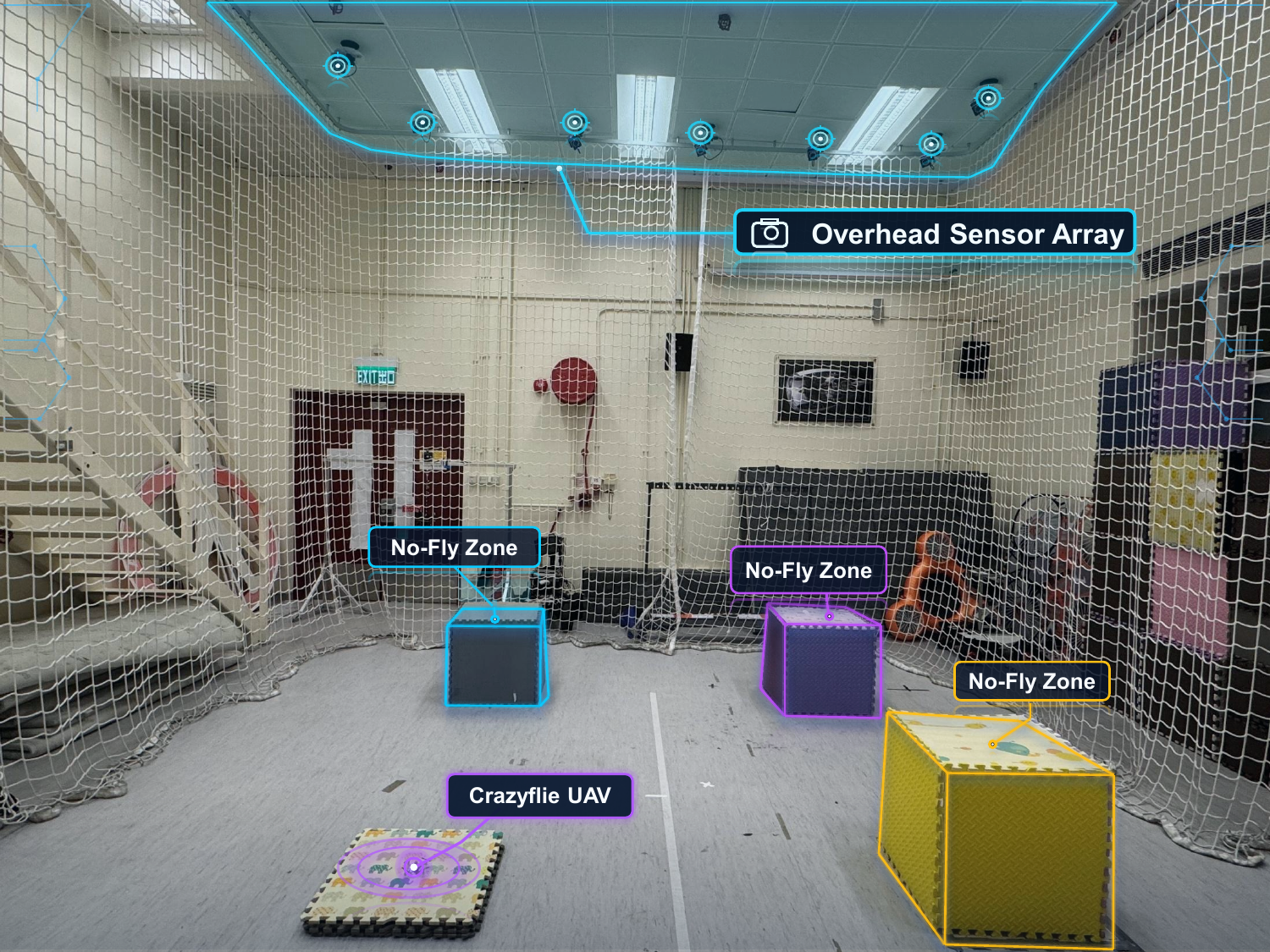}%
\caption{Crazyflie flight arena with designated no-fly zones.
The overhead Vicon system tracks the UAV for real-time intrusion-risk monitoring.}
\label{Fig_ExperimentalSetup}
\end{figure}

\begin{figure}[t]
\centering
\includegraphics[width=1.0\columnwidth]{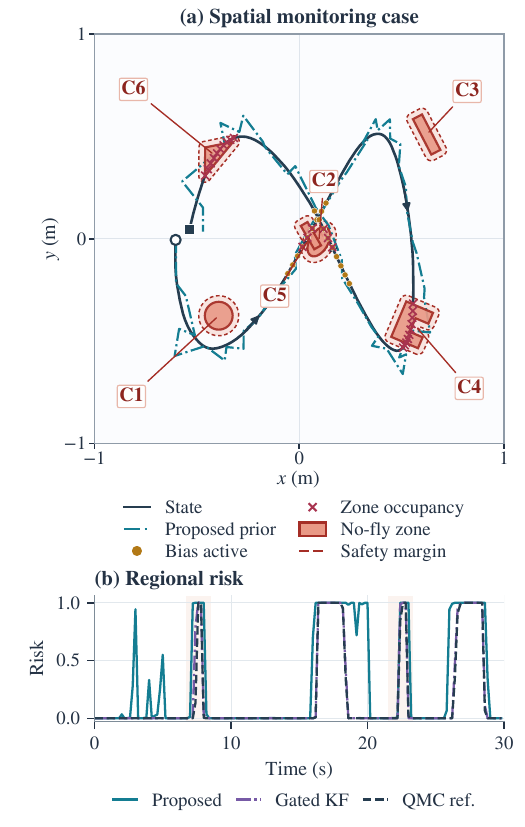}%
\caption{Spatial monitoring benchmark.
(a) State trajectory and the proposed monitor's trusted-prior mean.
An open circle marks the start and a filled square marks the end.
Dots indicate active bias, and crosses mark samples within a no-fly zone or its safety margin.
(b) Maximum regional risk for the proposed method, gated KF, and offline bias-known QMC reference.
Shading indicates intervals of active bias.}
\label{Fig_ApplicationCase}
\end{figure}

We evaluate the monitor during Crazyflie flights in the arena shown in Fig.~\ref{Fig_ExperimentalSetup}.
The overhead Vicon system tracks the quadrotor, and the monitor uses the position observations to assess the risk of entering each designated no-fly zone or its safety margin.
Test biases are applied only to the vulnerable observation channel.
The separate numerical benchmark in Fig.~\ref{Fig_ApplicationCase} uses state and bias references only for offline evaluation and computes reference intrusion probabilities by quasi-Monte Carlo (QMC) integration.
Comparative trials assess calibration and monitoring performance, while prescribed risk profiles test the temporal logic.

The estimator uses the state $\mathbf x_k=[\mathbf p_k^\top,\dot{\mathbf p}_k^\top]^\top$, observation matrices $\mathbf C=\mathbf H=[\mathbf I_2\ \mathbf0]$, and sampling interval $\Delta t=0.2$~s.
A constant-velocity model with white acceleration noise gives the estimator matrices
\begin{equation*}
\begin{aligned}
\mathbf A=
\begin{bmatrix}\mathbf I_2&\Delta t\mathbf I_2\\\mathbf0&\mathbf I_2\end{bmatrix},\
\mathbf Q=q_a
\begin{bmatrix}
\frac{\Delta t^3}{3}\mathbf I_2&\frac{\Delta t^2}{2}\mathbf I_2\\
\frac{\Delta t^2}{2}\mathbf I_2&\Delta t\mathbf I_2
\end{bmatrix}.
\end{aligned}
\end{equation*}
A fixed-grid search on 50 independent attack-free sequences gives $q_a=0.006$~m$^2$/s$^3$ for spatial monitoring.
The initial covariance is specified by
\[
\mathbf P_{0|-1}=\operatorname{diag}(0.004^2,0.004^2,0.006^2,0.006^2).
\]
The measurement noise covariances are $\mathbf R=0.006^2\mathbf I_2$ and $\mathbf R^t=\sigma_t^2\mathbf I_2$, with $\sigma_t=0.012$~m.
Trusted corrections follow risk assessment at 1~Hz ($m=5$).
All methods share observations and estimator parameters within each study.
We set $a_{\max}=0.018$~m, $(\alpha,\delta,\beta)=(0.01,0.005,0.005)$, and $(\tau^L,\tau^M,\tau^H)=(0.05,0.25,0.65)$.
Numerical certification uses $M=12$ and 8192 i.i.d. draws for the displayed case or 4096 for paired spatial comparisons.
The pointwise error budget under the calibrated model is $\delta+\beta=0.01$, with joint allocation as in \eqref{eq:numerical_total_failure}.

The trusted-only filter (``Trusted only'') reports occupancy under the trusted prior, whereas the nominal KF uses the zero-bias posterior in \eqref{eq:attack_conditioned_posterior}.
After an alarm, the gated KF switches to the trusted prior, while the alarm-only baseline (``Alarm only'' or ``Alarm-only bound'') switches from nominal to proposed risk.
The clipped KF scales innovations by $\min\{1,\sqrt{h_\alpha/b_k^a}\}$, retaining nominal covariance.
Bound only and No bound maximize over $\mathcal A_k$ and $\mathcal E_k^a(\delta)$, respectively, while the proposed method uses their intersection.
The witness estimate samples feasible biases, whereas the proposed bound adds cover inflation and upper probability endpoints.
The offline bias-known reference uses the benchmark motion model and trusted history, with analytic probabilities or independent scrambled Sobol QMC integration.

For reported risk $U$ and offline reference $p_k^\star$, a deficit occurs when $p_k^\star-U>10^{-12}$ and a material deficit when $p_k^\star-U>0.01$.
Excess $\mathbb E[(U-p_k^\star)_+]$ and shortfall $\mathbb E[(p_k^\star-U)_+]$ quantify conservatism and underestimation.
The joint miss and extra-high fractions are $\mathbb P(U\leq\tau,p_k^\star>\tau)$ and $\mathbb P(U>\tau,p_k^\star\leq\tau)$, respectively.
Their conditional rates use denominators $\mathbb P(p_k^\star>\tau)$ and $\mathbb P(p_k^\star\leq\tau)$, respectively.
For each bootstrap confidence interval (CI), resampling preserves trajectory-level pairing across methods, conditions, and times.

\subsection{Trajectory Monitoring and Statistical Calibration}

Fig.~\ref{Fig_ApplicationCase}(a) compares the trajectory and the proposed prior across six no-fly zones with $0.03$~m safety margins in an arena measuring $2$~m per side.
Biases point toward C2 and away from C5 during the test intervals.
The proposed envelope in Fig.~\ref{Fig_ApplicationCase}(b) covers the offline QMC reference at all 150 assessments.

\begin{figure}[t]
\centering
\includegraphics[width=1.0\columnwidth]{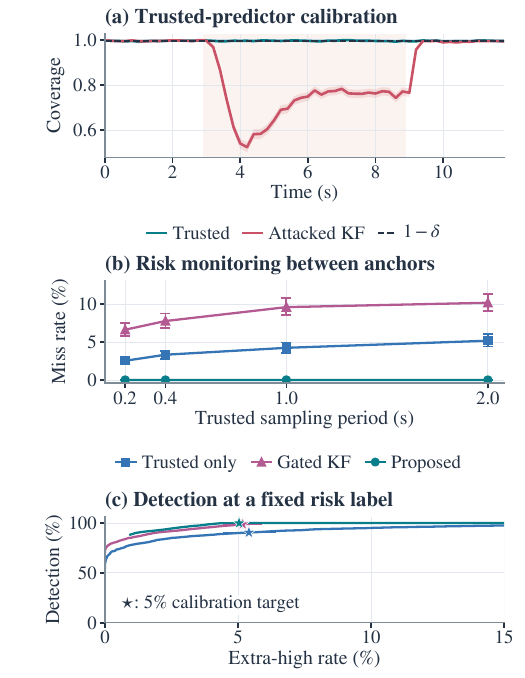}%
\caption{Calibration and monitoring performance.
(a) Coverage of the confidence set for the measurement bias.
Narrow bands show pointwise 95\% Wilson intervals, and background shading marks the attack interval.
(b) Conditional miss rates at $\tau=0.25$.
(c) Detection versus extra-high rate on 600 trajectories ($p_k^\star>0.25$ defines reference-high).
Stars mark operating points obtained with thresholds calibrated on 400 separate trajectories for a 5\% extra-high target.}
\label{Fig_StatisticalCalibration}
\end{figure}

The Gaussian studies use $q_a=6\times10^{-5}$~m$^2$/s$^3$ and an analytic half-space reference.
Calibration uses 2000 independent realizations of 60 updates under eight paired attack conditions.
In Fig.~\ref{Fig_StatisticalCalibration}(a), coverage ranges from 0.9910 to 0.9985 under trusted-only recursion but falls to 0.5255 when the filter feeds attacked observations back into its predictor in the strong concealment test.

Another 1000 paired realizations vary the trusted period over $\{0.2,0.4,1,2\}$~s under four attack policies.
As the period increases, Fig.~\ref{Fig_StatisticalCalibration}(b) shows rising conditional miss rates for Trusted only and the gated KF, while the proposed rate stays near zero.
At 1~Hz, Fig.~\ref{Fig_StatisticalCalibration}(c) gives detection rates of $99.84\%$ and $90.43\%$ for the proposed method and Trusted only, at conditional extra-high rates of $5.04\%$ and $5.41\%$, respectively.
The paired gain is $9.41$ percentage points (95\% CI $[7.92,11.12]$).

\begin{figure}[!t]
\centering
\includegraphics[width=1.0\columnwidth]{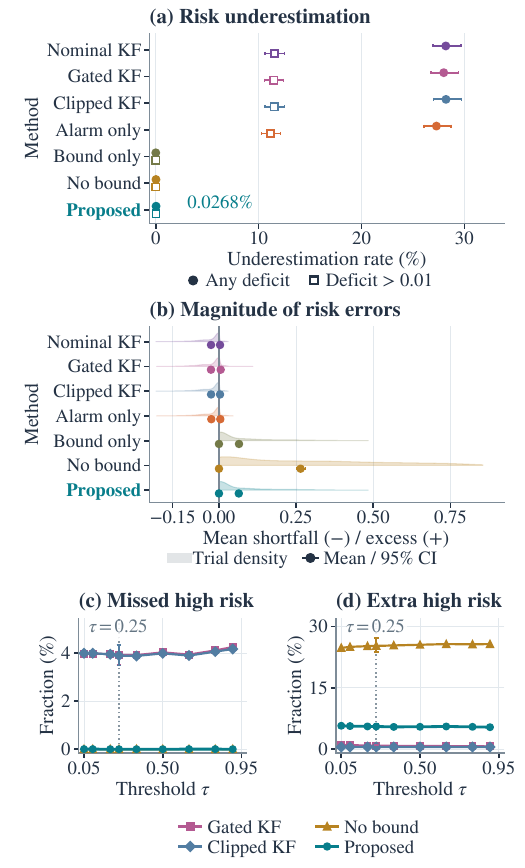}%
\caption{Comparative risk assessment.
(a) Any-deficit and material-deficit rates.
(b) Trajectory-mean shortfall (negative) and excess (positive).
(c) Joint miss fractions.
(d) Joint extra-high fractions.
Bars show paired 95\% confidence intervals, displayed only at $\tau=0.25$ in (c) and (d).}
\label{Fig_MethodComparison}
\end{figure}

\subsection{Comparative Performance and Component Ablations}

The comparison uses 1000 paired realizations across 100 conditions: five bias magnitudes from zero to $a_{\max}$, five trusted-noise levels from 4 to 30~mm, and four bias patterns (oblique concealment, rotation, trusted-prior tilt, and intermittent concealment).
The analytic half-space probability serves as the reference.
During the attack intervals, Fig.~\ref{Fig_MethodComparison}(a) gives a deficit rate of $0.0268\%$ for the proposed method (95\% CI $[0.0157,0.0423]\%$), versus $27.9961\%$, $28.2150\%$, and $27.2845\%$ for the gated KF, clipped KF, and Alarm only.
Fig.~\ref{Fig_MethodComparison}(b) shows small shortfalls for the proposed method, with mean excess $75.7\%$ lower than that of No bound and $1.0\%$ lower than that of Bound only.
At $\sigma_t=0.012$~m and peak bias $0.75a_{\max}$, the proposed method has fewer joint misses than the KF comparators in Fig.~\ref{Fig_MethodComparison}(c) and fewer extra-high decisions than No bound in Fig.~\ref{Fig_MethodComparison}(d).

\begin{figure}[!t]
\centering
\includegraphics[width=1.0\columnwidth]{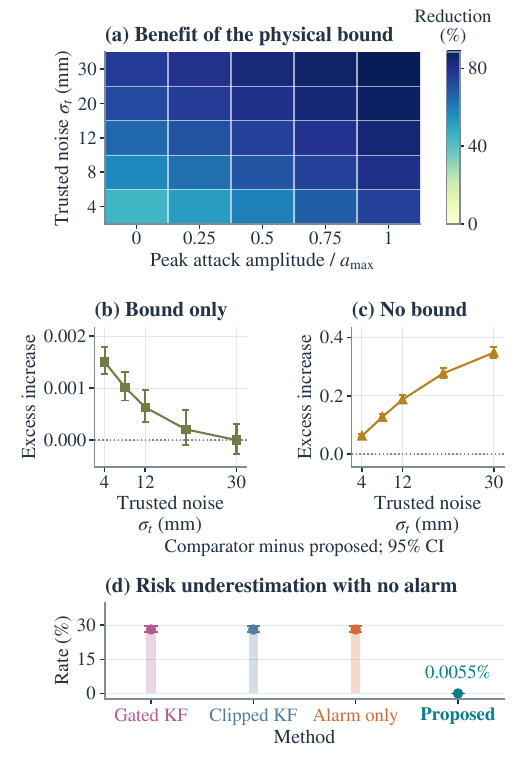}%
\caption{Component ablations.
(a) Reduction in mean excess relative to No bound versus normalized peak bias and trusted noise $\sigma_t$.
(b) Mean excess of Bound only minus that of the proposed method.
(c) Corresponding difference for No bound.
Positive differences favor the proposed method.
(d) Deficit rates conditional on no alarm.
Bars show paired 95\% confidence intervals.}
\label{Fig_MatchedAblations}
\end{figure}

For nonempty intersections, the exact proposed envelope cannot exceed the Bound only or No bound maximum.
Relative to No bound, the reduction in mean excess grows with trusted noise in Fig.~\ref{Fig_MatchedAblations}(a).
Fig.~\ref{Fig_MatchedAblations}(b) shows the largest benefit of the innovation constraint at lower noise, while Fig.~\ref{Fig_MatchedAblations}(c) shows the increasing benefit of the physical bound.
Among assessments with no alarm, Fig.~\ref{Fig_MatchedAblations}(d) gives deficit rates of $28.1464\%$ ($[26.9145,29.6072]\%$) for Alarm only and $0.0055\%$ ($[0.0022,0.0099]\%$) for the proposed method.
These conditional rates support retaining bias hypotheses after detector acceptance, whereas the pointwise guarantee precedes detector selection.

\subsection{Numerical and Spatiotemporal Validation}

\begin{figure}[!t]
\centering
\includegraphics[width=1.0\columnwidth]{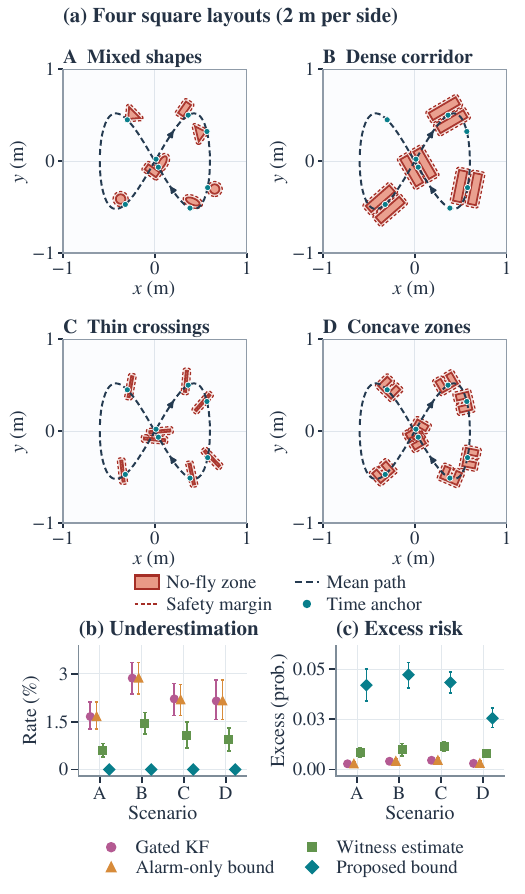}%
\caption{Four eight-region scenarios.
(a) No-fly zones, mean paths, and time markers in square layouts ($2$~m per side).
(b) Material-deficit rates.
(c) Mean excess over the bias-known reference.
Bars show paired 95\% confidence intervals.}
\label{Fig_MultiScenario}
\end{figure}

Analytic risk evaluation takes a median $0.040$~ms per sample excluding estimation.
Eight-region median and 95th-percentile monitoring times are $5.04$ and $12.20$~ms, respectively.
Layouts A--D in Fig.~\ref{Fig_MultiScenario}(a) pair mixed shapes with intermittent bias, dense corridors with nearest-center concealment, thin crossings with rotating bias, and concave zones with ramped alternating bias.
Each uses 24 paired sequences assessed at 16 indices with offline QMC references.
Fig.~\ref{Fig_MultiScenario}(b) shows no observed material deficits for the proposed method, versus pooled rates of $2.222\%$, $2.214\%$, and $1.009\%$ for the gated KF, Alarm only, and witness estimate.
The proposed method's mean excess in Fig.~\ref{Fig_MultiScenario}(c) is $0.0395$ across layouts.

\begin{figure}[!t]
\centering
\includegraphics[width=1.0\columnwidth]{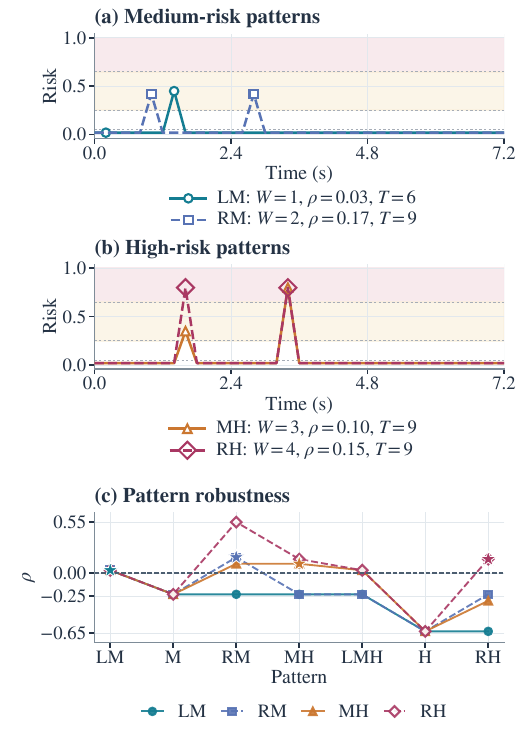}%
\caption{Temporal warnings.
(a) Isolated and repeated medium-risk events.
(b) Medium-to-high and repeated-high events.
In (a) and (b), horizontal lines mark risk thresholds, and shading identifies the medium- and high-risk ranges.
(c) Maximum robustness over each pattern's windows.
Stars mark explanations selected by severity and then robustness.
Legends report warning level $W$, margin $\rho$, and window $T$ in sampling increments ($\Delta t=0.2$~s).}
\label{Fig_TemporalExplanation}
\end{figure}

The windows $\mathcal T^q=\{3,6,9\}$ for $LM,RM,MH,RH$, $\{2,4,6\}$ for $M,H$, and $\{5,8,12\}$ for $LMH$ require 37 samples over $7.2$~s.
All profiles start at risk $0.02$.
Fig.~\ref{Fig_TemporalExplanation}(a) distinguishes isolated and repeated medium-risk events at warning levels 1 and 2.
The equal-peak profiles in Fig.~\ref{Fig_TemporalExplanation}(b) produce levels 3 and 4 for medium-to-high and repeated-high events.
Fig.~\ref{Fig_TemporalExplanation}(c) gives the selected robustness margins: $0.03$, $0.17$, $0.10$, and $0.15$.
Boolean, robustness, and incomplete-window checks verify the implementation.

\section{Conclusion}

This paper developed an intrusion-risk monitor for UAVs under bounded measurement attacks.
Trusted estimation, innovation statistics, and physical attack bounds identify admissible measurement biases even when the detector raises no alarm.
The monitor bounds regional occupancy probabilities over the resulting Gaussian posteriors, with confidence guarantees and finite-cover numerical certification.
PrSTL distinguishes ordered, persistent, and repeated risk and identifies each warning's severity, robustness margin, and supporting window.
Flight tests evaluated the monitor on a Crazyflie quadrotor in an arena with designated no-fly zones.
In controlled comparisons, the proposed method underestimated risk less often than the gated and clipped KFs and had lower mean excess than the variant without the physical bias bound.

\section*{Acknowledgment}

The authors would like to thank Ichiro Hasuo for his valuable guidance on formal methods.

\end{document}